\documentclass[a4paper,12pt]{article}
\usepackage[T2A]{fontenc}
\usepackage[utf8]{inputenc}   
\usepackage[russian,english]{babel}

\usepackage{verbatim} 	
\usepackage{amsmath, amsthm}
\usepackage{graphicx}
\usepackage{array} 	

\usepackage{rotating}	

\usepackage{hyperref}	

\renewcommand{\thefootnote}{\fnsymbol{footnote}}

\begin{document}
\pagestyle{plain}
\pagenumbering{arabic}

\selectlanguage{english}

\title{The radiative characteristics of quantum-well active region
of $AlGaAs$ lasers with separate-confinement heterostructure (SCH)}


\date{}

\author{S. I. Matyukhin, Z. Koziol\footnote{Corresponding author email: softquake@gmail.com}, and S. N. Romashyn,\\
Orel State Technical University,\\
29 Naugorskoye Shosse, Orel, 302020, Russia.}


\maketitle

\renewcommand{\thefootnote}{\arabic{footnote}}

\begin{abstract}

Computer simulations with Synopsys' Sentaurus TCAD 
are used to study the effect of the molar concentration of aluminum in 
the active and waveguide regions on the energy spectrum of carriers in Quantum Well (QW)
and the optical spectral characteristics of radiation of semiconductor lasers with $Al_x Ga_{1-x}As$ 
double heterostructures and separate confinement (SCH). Wavelength of single-mode lasers 
is shown to be almost independent of the concentration of aluminum 
in the waveguide, in a wide range of aluminum content, but to depend mainly on Al concentrations 
in QW region.

\end{abstract}

\baselineskip=3.00ex 




\section*{Introduction}

Currently, semiconductor lasers are used in CD/DVD readers/burners, 
telecommunication systems, precision measurement of displacement and high-resolution 
spectroscopy, for pumping solid state lasers, as well as in areas related to materials 
processing (cutting, microwelding) and medicine (laser surgery, dermatology). Since 1990, 
virtually all designs of lasers, in fact, are injection lasers with double 
heterostructure and separate confinement (SCH) \cite{Alferov}, \cite{Elisejev}.

In SCH lasers electronic transitions and photons generation are localized in the most narrow-band part 
- a quantum well (QW), whose width can reach several tens of angstroms, and an optical 
waveguide is a region of width about 1 $\mu m$. This separate confinement of electronic 
and optical excitations can significantly lower the threshold current density of laser generation 
and increase the power of semiconductor lasers operating in at room temperature.

Due to the relative simplicity and perfection of technology of the most commonly now 
used SCH lasers, solid solutions of $Al_x Ga_{1-x}As$ are used as wide-gap semiconductors. 
Reaching the threshold current density of these lasers less than $1 kA/cm^2$ at room 
temperature has opened up prospects for their practical application and served 
as a turning point in their production. However, further progress in this direction 
is associated with optimizing the design of laser diodes and, in particular, 
with the choice of the QW width and the values of the molar 
concentration $x$ of aluminum in the QW and waveguide regions.

In this paper, computer simulation of the influence 
of the molar concentration of aluminum in the QW and the waveguide region on the energy 
spectrum of charge carriers in the QW and on the spectral characteristics is performed. 
Wavelengths of a single-mode $Al_x Ga_{1-x}As$ lasers are calculated. It is shown that 
these wavelengths are almost independent of $Al$ concentration 
in the waveguide itself but are determined, primarily, by the concentration in the QW region.

The structure of SCH lasers we are studying consists of $Al_xGa_{1-x}As$ layer as an n-type emitter 
of thickness 1.5 $\mu m$, with doping concentration of $10^{18}/cm^3$, waveguide of $Al_xGa_{1-x}As$ of thickness 
$0.12 \mu m$ with n-type doping concentration of $5 \cdot 10^{15}/cm^3$,
an active laser region of $Al_xGa_{1-x}As$ of thickness $9, 12, 15$ or $18 nm$ with n-type doping concentration of $10^{15}/cm^3$,
again waveguide of $Al_xGa_{1-x}As$ of thickness $0.12 \mu m$ with n-type doping concentration of $5 \cdot 10^{15}/cm^3$, and 
a layer of $Al_xGa_{1-x}As$ as a p-type emitter of thickness 1.5 $\mu m$, with p-type doping concentration of $10^{18}/cm^3$.

An important aspect of using $Al_xGa_{1-x}As$ solid solutions is that this compound has two conduction-band 
minima. At $x<0.45$, a direct transition between conduction and valance bands dominates, while for higher $Al$
concentrations, an indirect transitions for a $E_c(k)$, with electron wave-vector $k$ shifted in 
$X$ direction, $<100>$, are energetically favorable. 
An approximation that is used in Sentaurus TCAD for $Al$ concentration
dependence of energy gap $E_g$ at $T=300 K$ has the form:

\begin{equation}\label{Eg}
\begin{array}{ll}
	E_g(x) =& 1.42248 + 0.56267 \cdot x, \; for \:  x<= 0.45,\\
	E_g(x) =& 1.98515 + 0.14835 \cdot (x-0.45) + 0.143 \cdot (x-0.45)^2, \; for \: x> 0.45
\end{array}
\end{equation}

\section*{The energy spectrum of carriers in the QW SCH lasers}

Emission of semiconductor SCH lasers is determined by interband transitions 
of electrons in the QW. Consequently, to determine the spectral characteristics 
of laser radiation it is important to know the energy spectrum 
of carriers in the QW.

Due to the specific structure of heterojunctions, this spectrum, as well as the wave 
functions describing the state of the carriers in the QW, can be found in the approximation 
of a rectangular potential well $V(y)$, on the basis of the stationary Schrödinger equation \cite{Coldren}:

\begin{equation}\label{schrodinger}
\left(
- \frac {\hbar ^2}{2} \frac {\partial}{\partial y} \left( \frac {1}{m(y)} \frac {\partial}{\partial y} 
+V(y) -E  + \frac {\hbar ^2 k ^2}{2 m(y)}  \right)
\right) \psi(y) = 0,
\end{equation}

where $k^2 = 2m(y)E/\hbar^2$, $\hbar$ is Planck's constant, $m(y)$ is effective mass 
(since the quantum well has different material properties than the waveguide, 
$m$ depends on the coordinate $y$, the axis assumed here to be directed perpendicular to the QW layer), and $E$
is energy of carriers relative to the bottom of quantum well. 

Eigenstates of \eqref{schrodinger} could be found analytically \cite{Coldren}. They are in the form of
symmetric and anti-symmetric stationary wave functions. The number of states in QW
is limited and depends on QW width $L$ and its potential depth, which we will call conduction band and valance band offsets, 
for electrons and holes, correspondingly, as well it depends on effective masses of charge carriers.

We choose to conduct simulation in a commercial software package Sentaurus TCAD 
of Synopsys \cite{tcad}, and to find energy of states from there. In practice, these energy levels
are found in Sentaurus' sdevice log files, which are plain text files and can be easy parsed by using 
custom written Perl\footnote{PERL is "Practical Extraction and Report Language", www.perl.org} scripts. 

We simulated operation of lasers with quantum well width equal to $9$, $12$, $15$, and $18 nm$, 
and various contents $x$ of aluminum in the active region and in the waveguide. It was assumed 
that the effective masses of electrons ($m$) and heavy ($m_h$) and light ($m_l$) holes, expressed in units 
of the rest mass of electron, varies linearly with molar 
concentration of alumina in $Al_x Ga_{1-x}As$ \cite{tcad}, and the temperature is $T=300K$. 
We have than, as a good approximation:

$m(x) = 0.057 \cdot x + 0.067$, 
$m_h(x) = 0.139 \cdot x + 0.481$, and 
$m_l(x) = 0.186 \cdot x + 0.074$.

\begin{figure}[h]
\begin{center}
      \resizebox{150mm}{!}{\includegraphics{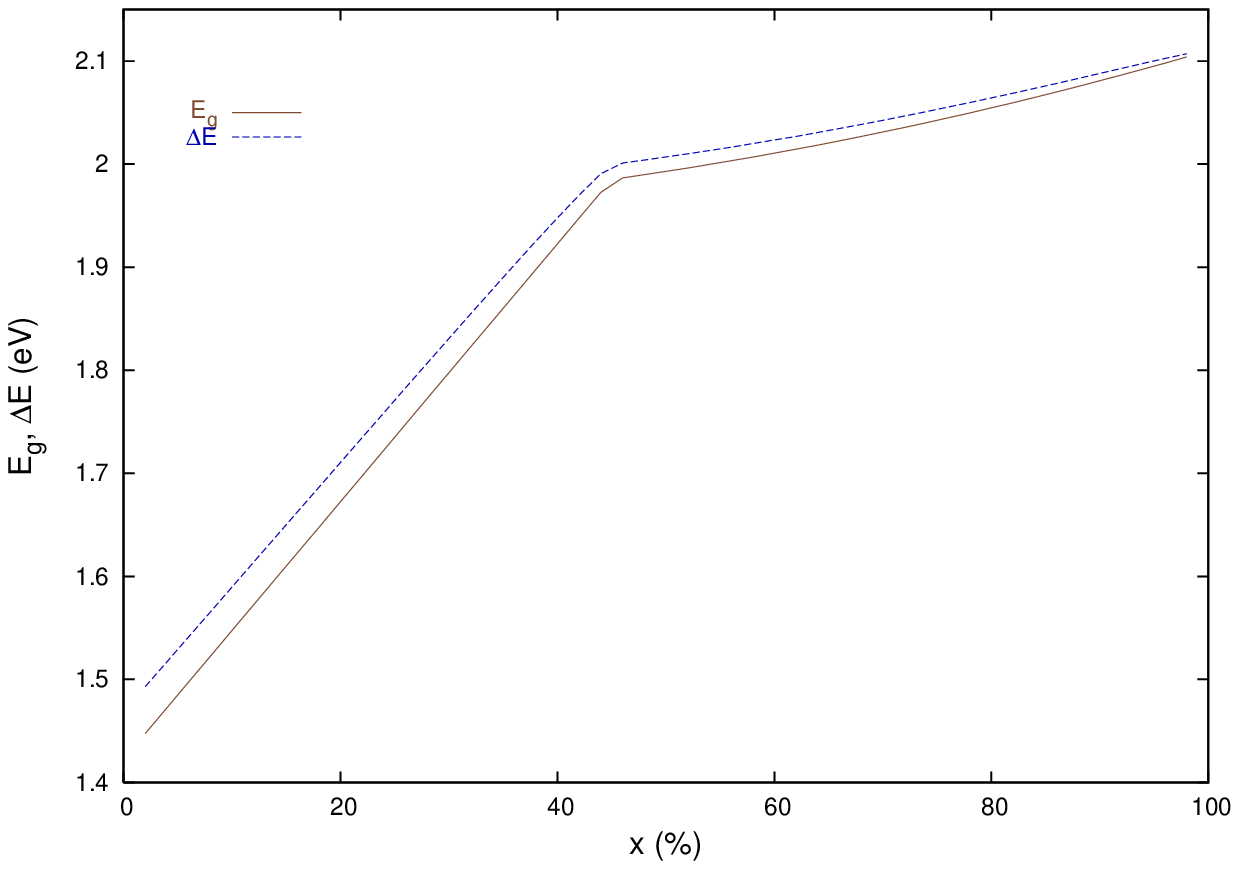}}
      \caption{Energy gap $E_g$ (when no quantum well effect is included) and energy difference between the first 
		conduction band state and the first valance band state in quantum well, as a function of $Al$ concentration in quantum well.
		It is assumed that $Al$ concentration in waweguide layers is $100\%$, i.e. that waveguide consists of $AlAs$.
		Calculations were done for quantum well of $9nm$ width.}
      \label{qwN_0030}
\end{center}
\end{figure}

The results of the simulation are presented in Figure \ref{qwN_0030}, which illustrates the difference 
between energy gap $E_g$ in QW, and $\Delta E$ which is given by $\Delta E=E_g + E_{c1} + E_{v1}$,
with $E_{c1}$ being the first conduction band electron state energy in QW, and $E_{v1}$ is
the first (closest to the top of valance band) energy level for heavy holes. 
An abrupt change in $E_g(x)$ and in $\Delta E(x)$, from a linear one to parabolic at $x=45\%$ is
explained by Eq. \eqref{Eg}; the main contribution to $\Delta E(x)$ is from $E_g(x)$ as well.

In next figures we show selected examples of more detailed results. 

\begin{figure}[t]
\begin{center}
      \resizebox{150mm}{!}{\includegraphics{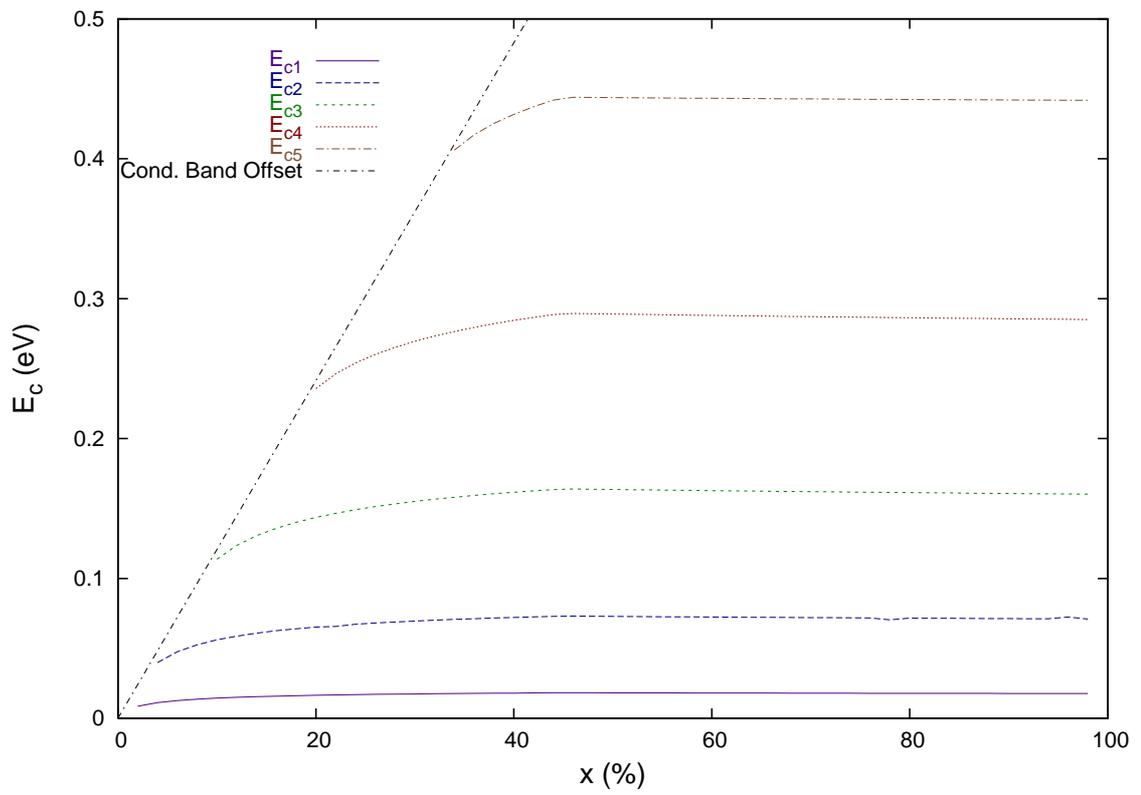}}
      \caption{Conduction band energy levels for a $15 nm$ width of QW.
		It was assumed that $Al$ concentration in QW is $0\%$ and $x$ is $Al$ concentration in waveguide.}
      \label{qwP_0003}
\end{center}
\end{figure}
\begin{figure}[t]
\begin{center}
      \resizebox{150mm}{!}{\includegraphics{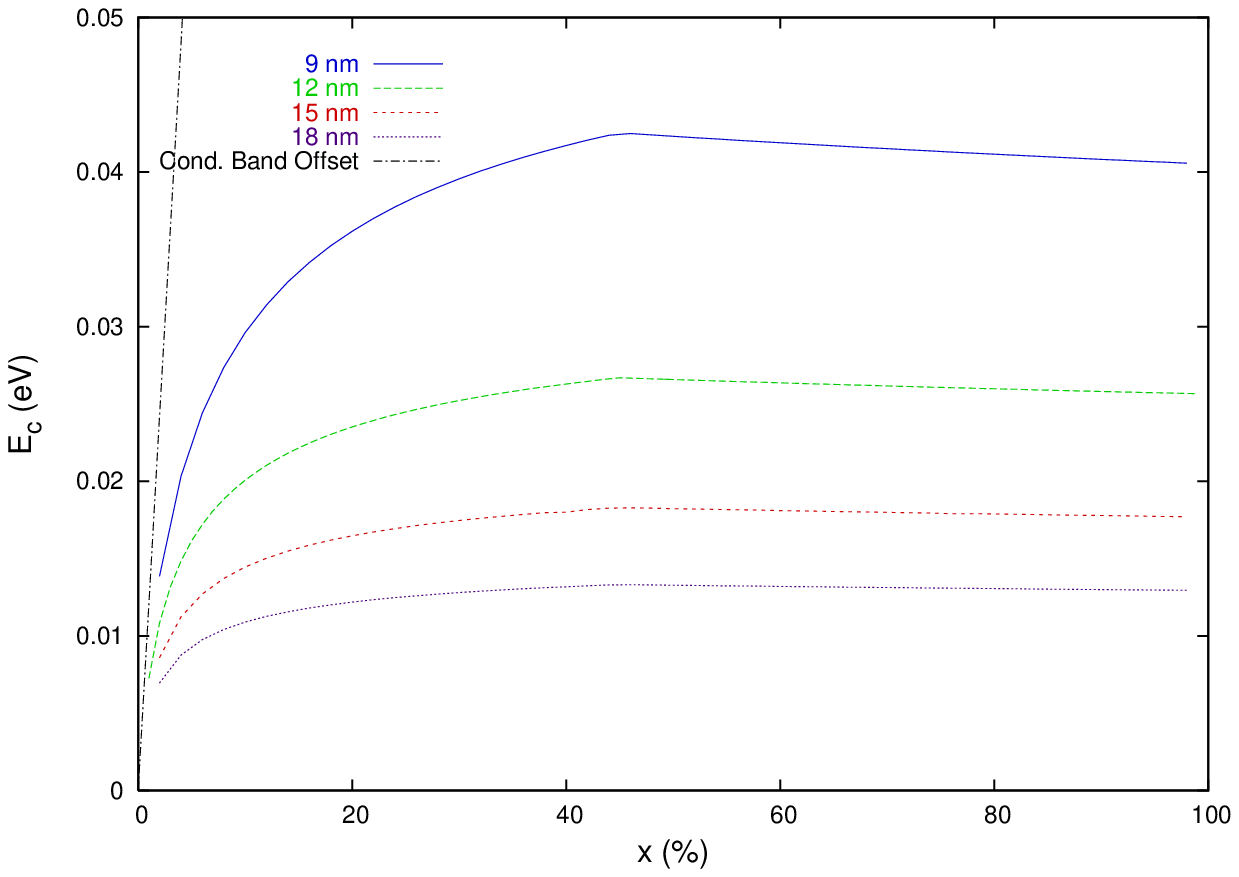}}
      \caption{Comparison of the first energy level in QW for electrons for various QW widths, $9$, $12$, $15$ and $18 nm$,
		It was assumed that $Al$ concentration in QW is $0\%$ and $x$ is $Al$ concentration in waveguide.}
      \label{qwP_0005}
\end{center}
\end{figure}

Hence, Figure \ref{qwP_0003} presents typical results for conduction band energy levels for a $15 nm$ width of QW,
as a function of $Al$ concentration in waveguide, while Figure \ref{qwP_0005} compares dependencies of the first
conduction band QW state energy $E_{c1}$ on $Al$ concentration, for various QW widths, $9$, $12$, $15$ and $18 nm$.

\begin{figure}[h]
\begin{center}
      \resizebox{150mm}{!}{\includegraphics{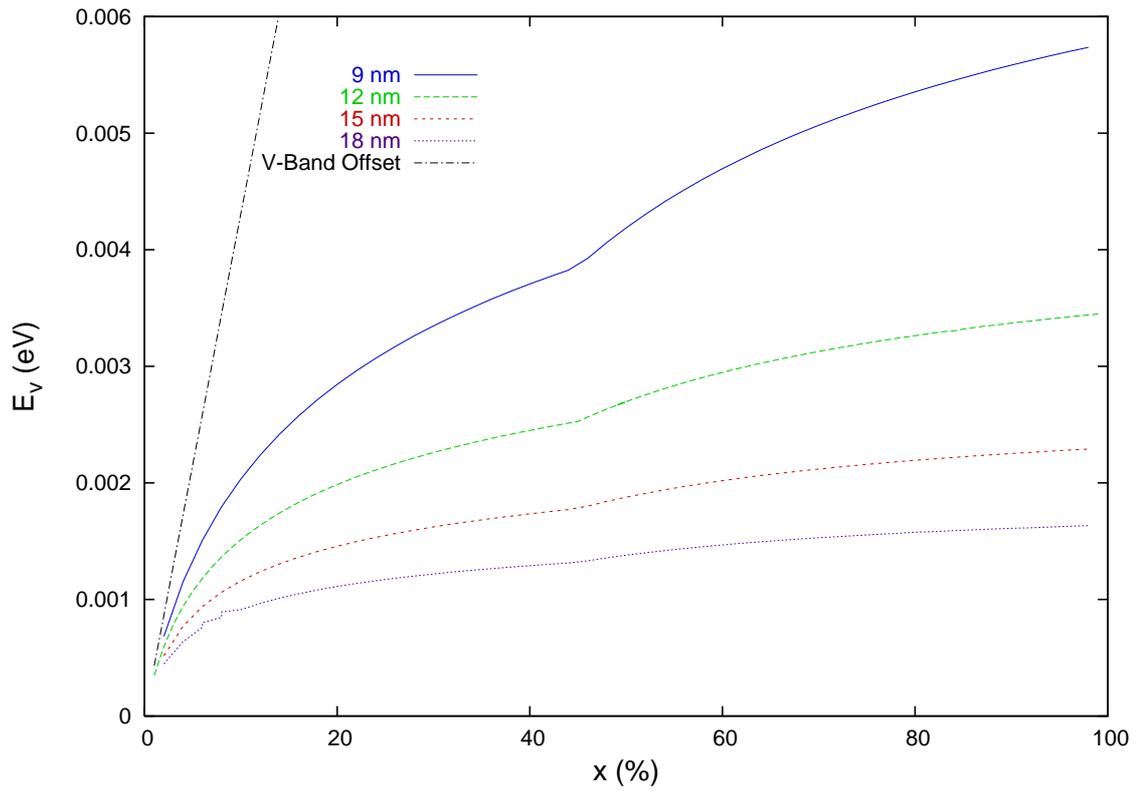}}
      \caption{Comparison of the first energy level in QW for heavy holes for various QW widths, $9$, $12$, $15$ and $18 nm$,
		It was assumed that $Al$ concentration in QW is $0\%$ and $x$ is $Al$ concentration in waveguide.
		Valance band offset
		is the difference between the valance band energy in the active region, and in waveguide, or, in other words, this is the depth of 
		quantum well for holes.
}
      \label{qwP_0006}
\end{center}
\end{figure}
\begin{figure}[h]
\begin{center}
      \resizebox{150mm}{!}{\includegraphics{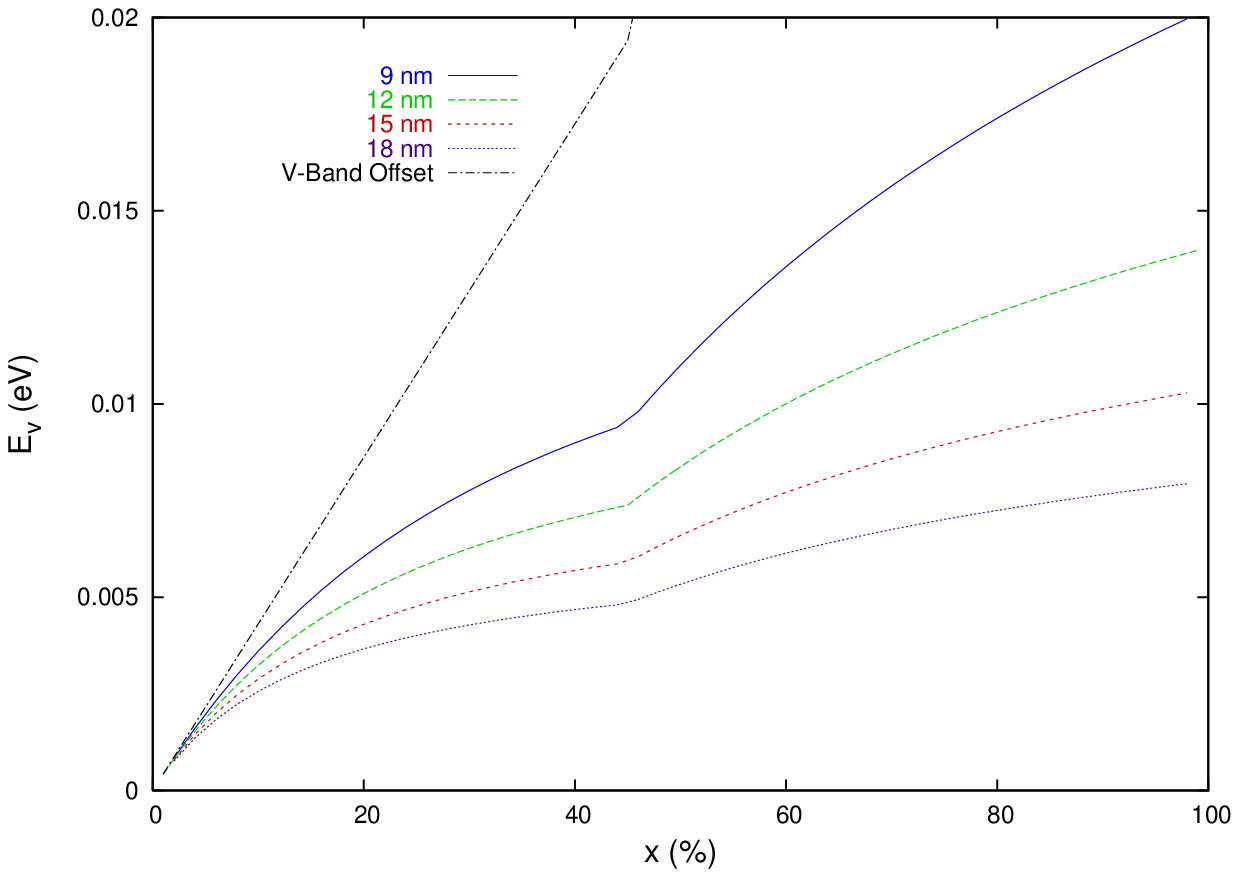}}
      \caption{Comparison of the first energy level in QW for light holes for various QW widths, $9$, $12$, $15$ and $18 nm$,
		It was assumed that $Al$ concentration in QW is $0\%$ and $x$ is $Al$ concentration in waveguide.}
      \label{qwP_0007}
\end{center}
\end{figure}

Figure \ref{qwP_0006} compares dependencies of the first valance band QW energy state of heavy holes, $E_{hh1}$, 
on $Al$ concentration, for various QW widths, $9$, $12$, $15$ and $18 nm$,
while Figure \ref{qwP_0007} shows similar results for light holes.

\begin{figure}[h]
\begin{center}
      \resizebox{150mm}{!}{\includegraphics{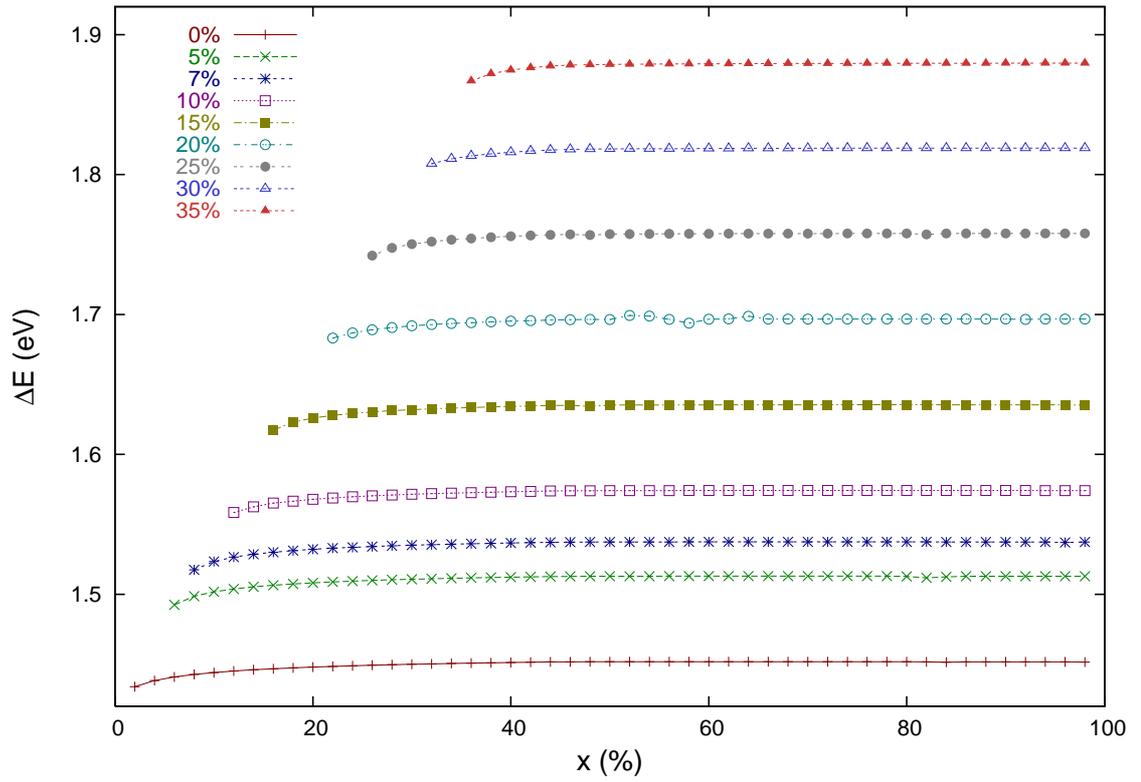}}
      \caption{Transition energy between the first electron subband and the first heavy holes subband in QW,
		for $12nm$ width of QW active region. $x$ is concentration of $Al$ in waveguide, while every curve 
		is computed for different values of $Al$ concentration in active region of QW, as indicated in the figure.}
      \label{swap_0060}
\end{center}
\end{figure}

We performed also calculations for dependence of QW energy levels on $Al$ concentration 
in waveguide, for a broad range of initial $Al$ concentrations in QW, for all studied QW widths.
A typical example data, for $12 nm$ width of QW, are shown in Figure \ref{swap_0060}.

\begin{figure}[h]
\begin{center}
      \resizebox{150mm}{!}{\includegraphics{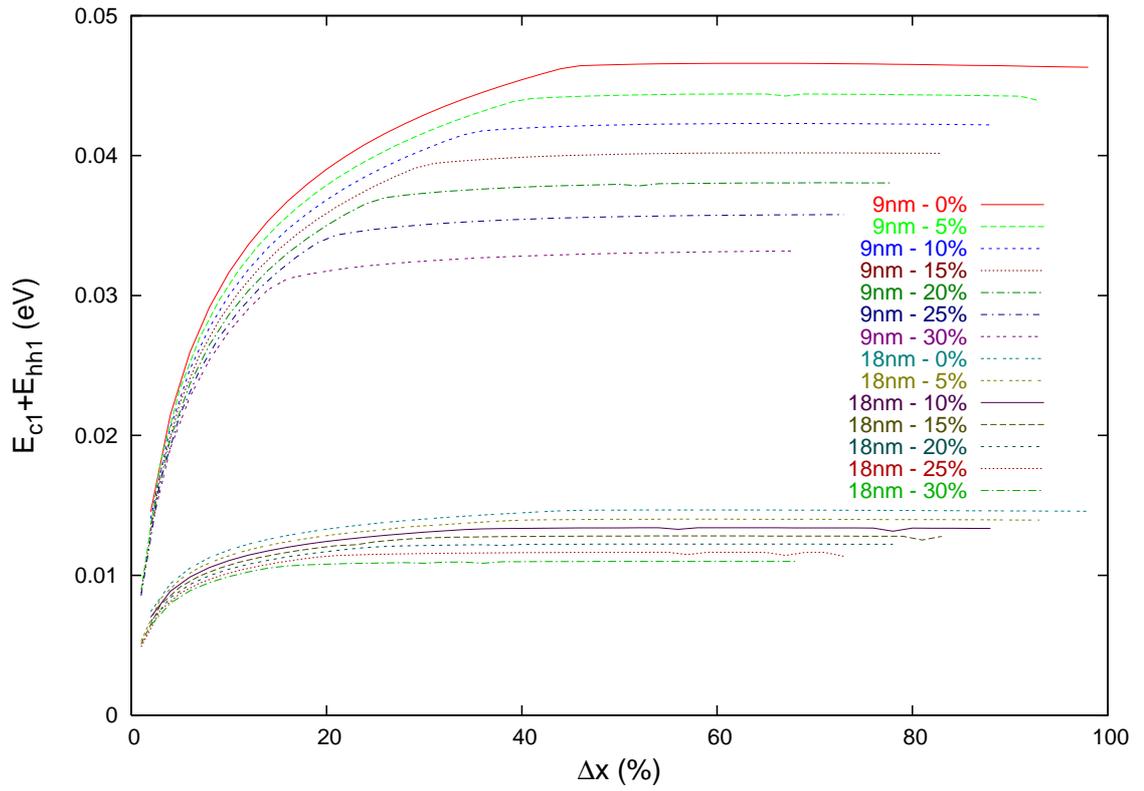}}
      \caption{Comparison of QW states energy $E_{c1}+E_{hh1}$ as a function of the difference 
		between $Al$ concentration in waveguide and in the active region, $\Delta x$, 
		for the case of $9nm$ and $18nm$ QW width, and different $Al$ concentrations in the active regions,
		as indicated in the figure.}
      \label{swap_0070}
\end{center}
\end{figure}

A question arises, to what an extend the contribution of energy of QW states depends 
on the difference of $Al$ concentrations between waveguide and the active region. To analyze that, 
we draw Figure \ref{swap_0070}. There, $E_{c1}+E_{hh1}$ is shown as a function of the difference 
between $Al$ concentration in waveguide and in active region, $\Delta x$, for the case of $9nm$ and $18nm$ 
QW width, and different $Al$ concentrations in the active region. The most interesting region, from 
the engineering point of view, is below $\Delta x$ of around $30\%$. We see that in that region, 
while at lower QW width of $9nm$ differences between curves reach energies of around $10 meV$, for QW width of
$18nm$ they become less than about $5 meV$.


\section*{Radiation characteristics of $AlGaAs$-based SCH lasers} 

Spectral characteristics of laser radiation are determined by the energy difference 
between the quantization levels of electrons and holes in the QW, 
as well as the characteristics of the resonator, which is used to enhance radiation. 
The maximum wavelength of the radiation produced in the active region of the laser 
diode may be calculated by the formula:

\begin{equation}\label{lambda}
\lambda = \frac {hc}{E_g + E_{c1} + E_{hh1}},
\end{equation}

where $h$ is Planck's constant and $c$ - the speed of light. 

As shown by simulation results, 
the energy states of electrons, $E_{c1}$, and holes, $E_{hh1}$, change very slowly with 
the changes in concentration difference of aluminum in the active region 
and waveguide. 

Moreover, for a broad range of $Al$ concentrations, the energy contribution 
from QW states does nearly not depend on the concentration of 
aluminum itself in the waveguide but is determined rather by $Al$ concentration diffrences 
between the active region (QW) and waveguide.

\begin{figure}[h]
\begin{center}
      \resizebox{150mm}{!}{\includegraphics{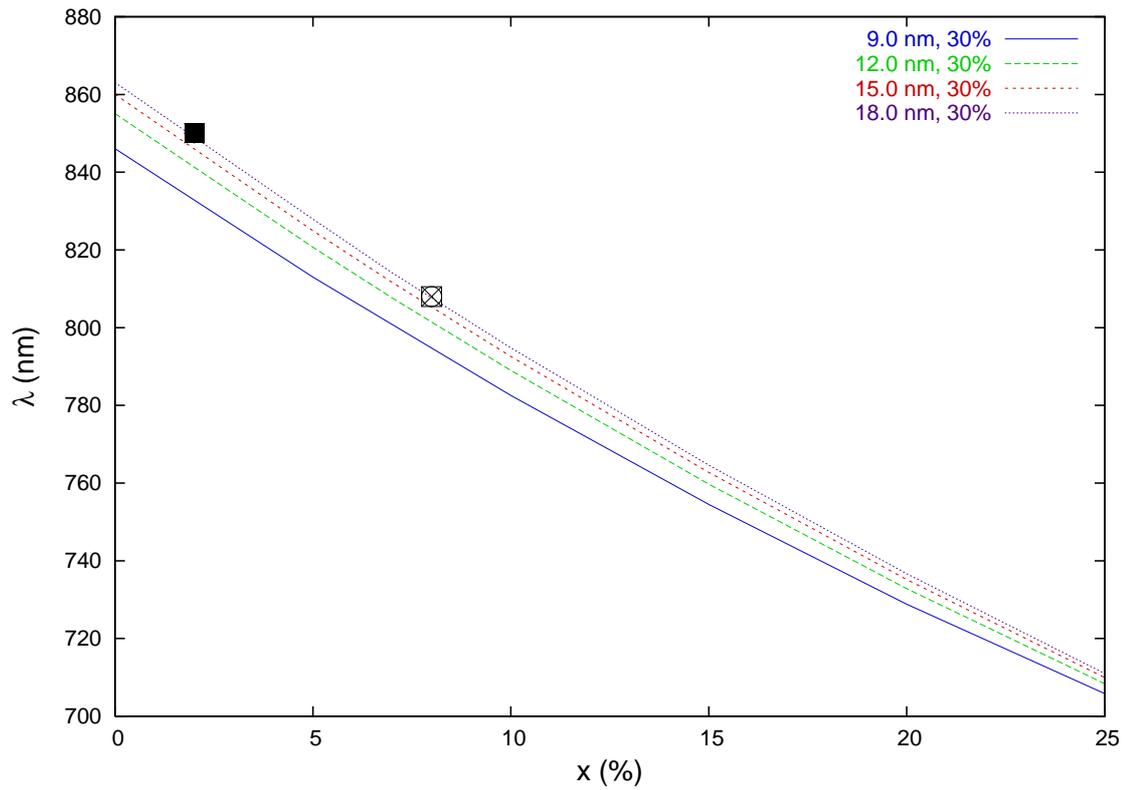}}
      \caption{Lasing wavelength as a function of $Al$ concentration in active region, when waveguide concentration is 
		$30\%$ of $Al$, for several values of QW width, as indicated in the figure.
		Markers correspond to wavelengths of high-power semiconductor SCH lasers produced 
		at Research Institute "Polyus" in Moscow.}
      \label{lambda30}
\end{center}
\end{figure}

The results of calculations of the wavelength $\lambda$, 
for SCH lasers with QW width equal to $9$, $12$, $15$, and $18 nm$, are presented in Figure \ref{lambda30}.
Markers in this figure correspond to wavelengths of high-power 
semiconductor lasers \cite{Andrejev_1}, and \cite{Andrejev_2}, 
produced in the Research Institute "Polyus" in Moscow.

We ought to emphasize that we do not expect the existence of lasing action when $Al$ concentration
exceeds $45\%$, as electronic transitions would require then a change of wavevector $\Delta k>0$,
and these processes would require interaction with phonons. For that reason, however, studing 
physical processes around these $Al$ concentrations is extremely interesting.


\section*{Conclusions}

The depth of the QW, and the energy of the 
electrons and holes in the active region of $Al_x Ga_{1-x}As$ lasers with a given 
QW width, vary with the difference of the molar concentration of aluminum in the active 
region and waveguide regions. 

In this single-mode lasers the lasing maximum wavelength $\lambda$
is determined by the semiconductor bandgap in the QW, as well as the lowest energy states 
of electrons and holes in the QW.

At low $Al$ concentrations in waveguide, energy levels in QW for both electrons and holes
increases monotonically with concentration of aluminum, while at higher $Al$ concentrations in waveguide, 
of around $50\%$ and more, they become independent on concentration of aluminum.

The lasing wavelength does nearly not depend on the concentration of aluminum in the 
waveguide and is determined mainly by 1) the concentration in the active region of laser 
diodes and 2) by the difference between $Al$ concentrations in waveguide and active regions. 

With the increase of the concentration in active region, the lasing wavelength $\lambda$ 
decreases, in agreement with results obtained at the Research Institute "Polyus" in Moscow.

These results are important from the point of production of SCH lasers based 
on solid solutions of $Al_x Ga_{1-x}As$ and can be used to optimize the design of these 
lasers to improve the efficiency of converting electrical energy into coherent 
laser radiation.

The work was carried out under the Federal Program "Research and scientific-pedagogical 
cadres of Innovative Russia" (GC number P2514).


\clearpage

\section*{Figure Captions}

\end{document}